\documentclass[]{spie}  %>>> use for US letter paper
\usepackage[]{graphicx}
\usepackage{amsfonts,amsmath,amssymb}
\usepackage{eucal}
\def\openone{\leavevmode\hbox{\small1\kern-3.3pt\normalsize1}}

\def\bbbz{{\Bbb Z}}
\def\ad{\mbox{ad\,}}

\def\diag{\mbox{diag\,}}
\def\fr#1{{\mathfrak{#1}}}

\def\openone{\leavevmode\hbox{\small1\kern-3.3pt\normalsize1}}

\def\m{{\mathbf m}}

\def\ad{\mathrm{ad\,}}

\def\bdiag{\mbox{block-diag\,}}
\def\diag{\mbox{diag\,}}

\def\bbbe\mathbb{E}

\def\bbbc{\mathbb{C}}
\def\bbbe{\mathbb{E}}
\def\bbbz{\mathbb{Z}}

\def\ad{\mbox{ad\,}}

\title{Bose-Einstein condensates with $F=1$ and $F=2$.
Reductions and soliton interactions of multi-component NLS models.}

\author{V. S. Gerdjikov\supit{}, N. A. Kostov\supit{} and T. I. Valchev\supit{}
\skiplinehalf
\supit{}Institute for Nuclear Research and Nuclear Energy, \\
Bulgarian academy of sciences\\ 72 Tsarigradsko chaussee, 1784
Sofia, Bulgaria
}

\authorinfo{Further author information: (Send correspondence to V. S. Gerdjikov)\\ V. S. Gerdjikov:
E-mail: gerjikov@inrne.bas.bg, Telephone: +3592 979 5638 }
\begin{document}
  \maketitle

\begin{abstract}
We analyze  a class of multicomponent nonlinear Schr\"odinger
equations (MNLS) related to the symmetric {\bf BD.I}-type symmetric spaces and their
reductions. We briefly outline the direct and the inverse scattering method for
the relevant Lax operators and the soliton solutions. We use the Zakharov-Shabat dressing method to obtain
the two-soliton solution and analyze the soliton interactions of the MNLS equations and some of their reductions.

\end{abstract}

%>>>> Include a list of keywords after the abstract

\keywords{ Bose-Einstein condensates, Multicomponent nonlinear Schr\"odinger equations,   Soliton solutions,
Soliton interactions}

%%%%%%%%%%%%%%%%%%%%%%%%%%%%%%%%%%%%%%%%%%%%%%%%%%%%%%%%%%%%%
\section{INTRODUCTION}
\label{sec:intro}  % \label{} allows reference to this section

Bose-Einstein condensate (BEC) of alkali atoms in the $F=1$ hyperfine state,
elongated in $x$ direction and confined in the transverse
directions $y,z$ by purely optical means are described by a  3-component
normalized spinor wave vector  $ {\bf\Phi}(x,t)=(\Phi_1, \Phi_0 , \Phi_{-1})^{T}(x,t)$ satisfying the
nonlinear Schr\"{o}dinger (MNLS) equation \cite{IMW04} see also \cite{imww04,LLMML05,uiw06,dwy08,Kevre*08}:
\begin{eqnarray}\label{eq:1}
&& i\partial_{t} \Phi_{1}+\partial^{2}_{x} \Phi_{1}+2(|\Phi_{1}|^2
+2|\Phi_{0}|^2) \Phi_{1} +2\Phi_{-1}^{*}\Phi_{0}^2=0, \nonumber \\
&& i\partial_{t} \Phi_{0}+\partial^{2}_{x} \Phi_{0}+2(|\Phi_{-1}|^2
+|\Phi_{0}|^2+|\Phi_{1}|^2) \Phi_{0} +2\Phi_{0}^{*}\Phi_{1}\Phi_{-1}=0,\\
&& i\partial_{t}\Phi_{-1}+\partial^{2}_{x}
\Phi_{-1}+2(|\Phi_{-1}|^2+ 2|\Phi_{0}|^2) \Phi_{-1}
+2\Phi_{1}^{*}\Phi_{0}^2=0. \nonumber
\end{eqnarray}
spinor BEC with $F=2$ for rather specific choices of the scattering lengths in dimensionless
coordinates takes the form \cite{uiw07}:
\begin{eqnarray}
&&i\partial_t\Phi_{\pm 2}+\partial_{xx}\Phi_{\pm 2} +2 (\vec{{\bf \Phi}},\vec{{\bf \Phi^{*}}}) \Phi_{\pm 2} -
(2\Phi_{2}\Phi_{-2}-2 \Phi_{1}\Phi_{-1}+\Phi_{0}^2) \Phi_{\mp 2}^*,\nonumber \\
&&i\partial_t\Phi_{\pm 1}+\partial_{xx}\Phi_{\pm 1} +2 (\vec{{\bf \Phi}},\vec{{\bf \Phi^{*}}}) \Phi_{\pm 1} -
(2\Phi_{2}\Phi_{-2}-2 \Phi_{1}\Phi_{-1}+\Phi_{0}^2) \Phi_{\mp 1}^*,  \\
&&i\partial_t\Phi_{0}+\partial_{xx}\Phi_{0} +2 (\vec{{\bf \Phi}},\vec{{\bf \Phi^{*}}}) \Phi_{0} -
(2\Phi_{2}\Phi_{-2}-2 \Phi_{1}\Phi_{-1}+\Phi_{0}^2) \Phi_{0}^*.\nonumber
\end{eqnarray}

Both models have natural Lie algebraic interpretation and are related to the symmetric
spaces ${\bf BD.I}\simeq {\rm SO(n+2)}/{\rm SO(n)\times SO(2)}$ with $n=3$ and $n=5$ respectively.
They are integrable by means of inverse scattering transform
method \cite{ForKu*83,tw98,GKV*09}. Using a modification of the Zakharov-Shabat
`dressing method' we describe the  soliton solutions \cite{IMW04,kagg07} and the effects of the reductions on them.

Sections 2  and 3 contain the basic details on the direct and inverse scattering problems for the Lax
operator. Section 4 outlines the effects of the algebraic reductions of the MNLS. In Section 5 using the
Zakharov-Shabat dressing method we derive the one- and two-soliton solutions of the MNLS and discuss their
 properties. Section 6  is dedicated to the analysis of the soliton interactions of the MNLS.
 To this end we evaluate the limits of the generic  two-soliton solution
 for $t\to\pm\infty$. As a result we establish that the effect of the interactions on the soliton
 parameters is analogous to the one for the scalar NLS equation and consists in shifts of the `center of mass'
 and shift in the phase.

%`dressing method' we describe the  \\ soliton solutions \cite{IMW04,kagg07} and the Hamiltonian properties
%\\ of these equations.

%\newpage

\section{The method for solving MNLS with $F=1$ and $F=2$}

MNLS equations  for the {\bf BD.I.} series of symmetric spaces
(algebras of the type $so(n+2)$ and $J$ dual to $e_1$) have the
Lax representation $[L,M]=0$ as follows \cite{ForKu*83,GKKV*08,GKV*09}
\begin{equation}\label{eq:3.1}
\begin{aligned}
L\psi (x,t,\lambda ) &\equiv  i\partial_x\psi + U(x,t,\lambda)\psi  (x,t,\lambda )=0, &\qquad
M\psi (x,t,\lambda ) &\equiv  i\partial_t\psi + V(x,t,\lambda)\psi  (x,t,\lambda )=0,\\
U(x,t,\lambda) &= Q(x,t) - \lambda J , &\qquad    V(x,t,\lambda) &= V_0(x,t) +
\lambda V_1(x,t) - \lambda ^2 J, \\
V_1(x,t)&= Q(x,t), &\qquad V_0(x,t) &= i \ad_J^{-1} \frac{d Q}{dx}
+ \frac{1}{2} \left[\ad_J^{-1} Q, Q(x,t) \right].
\end{aligned}
\end{equation}
where $\ad_J X=[J,X]$ and $\ad_J^{-1}$ is well defined  on the image of $\ad_J$ in $\fr{g}$;
\begin{equation}\label{vec1}
Q=\left(\begin{array}{ccc}  0 & \vec{q}^{T} & 0 \\
  \vec{p}{\,}^* & 0 & s_{0}\vec{q} \\  0 & \vec{p}{\,}^\dag s_{0} & 0 \\
\end{array}\right),\qquad J=\mbox{diag}(1,0,\ldots 0, -1).
\end{equation}
The  vector $\vec{q}$ for $F=1$ (resp. $F=2$) is $3$- (resp. $5$-) component and has the form
\begin{equation}\label{eq:sok}
\vec{q} = (\Phi_1,\Phi_0 , \Phi_{-1})^T, \qquad \vec{q} = (\Phi_{2}, \Phi_1,\Phi_0 , \Phi_{-1} , \Phi_{-2})^T,
\end{equation}
and the corresponding matrices $s_0  $ enter in the definition of $so(2r+1)$ with $r=2$ and $r=3$:
\begin{equation}\label{eq:z1.6a}
 X\in so(2r+1), \qquad X + S_0 X^T S_0 =0, \qquad S_0  =
 \sum_{s=1}^{2r+1} (-1)^{s+1} E_{s, n+1-s}, \qquad S_0 = \left(\begin{array}{ccc}  0 & 0 & 1 \\
0 & -s_0 & 0 \\  1 & 0 & 0 \\ \end{array}\right).
\end{equation}
By $E_{sp}$ above we mean $2r+1\times 2r+1$ matrix with matrix
elements $(E_{sp})_{ij}=\delta_{si}\delta_{pj}$.
With the definition of orthogonality used in (\ref{eq:z1.6a}) the Cartan generators $H_k=E_{k,k}
-E _{n+3-k,n+3-k}$ are represented by diagonal matrices.

If we make use of the typical reduction $Q=Q^{\dag}$ (or $\vec{p}{\,}^* =\vec{q}$)
the generic MNLS type equations related to ${\bf BD.I.}$ acquire the form:
\begin{equation}\label{eq:4.2}
i \vec{q}_t+ \vec{q}_{xx} + 2 (\vec{q}{\,}^\dag,\vec{q}) \vec{q} -
(\vec{q},s_0\vec{q}) s_0\vec{q}{\,}^* =0.
\end{equation}
The Hamiltonians for the MNLS equations (\ref{eq:4.2})  are given by
\begin{eqnarray}\label{eq:Ham1}
H_{{\rm MNLS}}=\int_{-\infty}^\infty d x
\left((\partial_{x}\vec{q}{\,}^\dag,\partial_{x}\vec{q})- (\vec{q}{\,}^\dag,\vec{q})^2+ \frac{1}{2}
| (\vec{q}^T,s_{0}\vec{q})|^2\right).
\end{eqnarray}

\section{The Direct and the Inverse scattering problem }\label{sec:3}

\subsection{The fundamental analytic solution}

We remind some basic features of the inverse scattering theory for the
Lax operators $L$, see  \cite{GKKV*08,GKV*09}. There we have made use of the general theory
developed in \cite{ZaSh1,ZaSh2,ZMNP,FaTa, VSG1} and the references therein.
The Jost solutions  of $L$ are defined by:
\begin{equation}
\lim_{x \to -\infty} \phi(x,t,\lambda) e^{  i \lambda J x
}=\openone, \qquad  \lim_{x \to \infty}\psi(x,t,\lambda) e^{  i
\lambda J x } = \openone
\end{equation}
and the scattering matrix $T(\lambda,t)\equiv
\psi^{-1}\phi(x,t,\lambda)$. The special choice of $J$ and the
fact that the Jost solutions and the scattering matrix take values
in the group $SO(2r+1)$ we can use the following block-matrix
structure of $T(\lambda,t)$
\begin{equation}\label{eq:25.1}
T(\lambda,t) = \left( \begin{array}{ccc} m_1^+ & -\vec{b}^-{}^T & c_1^- \\
\vec{b}^+ & {\bf T}_{22} & - s_0\vec{B}^- \\ c_1^+ & \vec{B}^+{}^Ts_0 & m_1^- \\
\end{array}\right), \qquad \hat{T}(\lambda,t) = \left( \begin{array}{ccc} m_1^- & \vec{B}^-{}^T & c_1^- \\
-\vec{B}^+ & {\bf \hat{T}}_{22} &  s_0\vec{b}^- \\ c_1^+ & -\vec{b}^+{}^Ts_0 & m_1^+ \\
\end{array}\right),
\end{equation}
where $\vec{b}^\pm (\lambda,t)$ and $\vec{B}^\pm (\lambda,t)$ are
$2r-1$-component vectors, ${\bf T}_{22}(\lambda)$ is $2r-1 \times 2r-1$ block matrix, and $m_1^\pm
(\lambda)$, and $c_1^\pm (\lambda)$ are scalar functions. Such
parametrization is compatible with the generalized Gauss
decompositions of $T(\lambda)$ which read  as follows:
\begin{equation}\label{eq:25.1'}
\begin{aligned}
T(\lambda,t) &= T^-_J D^+_J \hat{S}^+_J , &\quad  T(\lambda,t) &= T^+_J D^-_J \hat{S}^-_J ,\\
T^\mp_J &= e^{\pm \left(\vec{\rho}^\pm , \vec{E}^\mp_1 \right)}  , &
S^\pm _J &= e^{\pm \left(\vec{\tau}^\pm , \vec{E}^\pm_1 \right)} ,  & \qquad
D_J^\pm &= \diag \left( (m_1^\pm)^{\pm 1} , {\bf m}_2^\pm , (m_1^\pm)^{\mp1} \right),
\end{aligned}
\end{equation}
where
\begin{equation}\label{eq:25.1''}
\begin{aligned}
\left(\vec{\rho}^+ , \vec{E}^+_1 \right) &= \sum_{k=1}^{r-1} (\rho^+_{k} E_{e_1-e_{k+1}} + \rho^+_{\bar{k}}
E_{e_1+e_{k+1}}) + \rho^+_{r} E_{e_1}, \\  \left(\vec{\rho}^- , \vec{E}^-_1 \right) &=
\sum_{k=1}^{r-1} (\rho^-_{k} E_{-e_1+e_{k+1}} + \rho^-_{\bar{k}} E_{-e_1-e_{k+1}}) + \rho^-_{r} E_{-e_1},
\end{aligned}
\end{equation}
and similar expressions for $\left(\vec{\tau}^\pm , \vec{E}^\pm_1 \right)$.
The functions $m_1^\pm$ and $n\times n$ matrix-valued functions ${\bf m}_2^\pm$
are analytic for $\lambda\in\bbbc_\pm$. We have introduced also the notations:
\begin{equation*}\label{eq:25.1a}
%\begin{aligned}
\vec{\rho}^- =\frac{\vec{B}^-}{m_1^-},  \qquad \vec{\tau}^-
=\frac{\vec{B}^+}{m_1^-}, \qquad \vec{\rho}^+
=\frac{\vec{b}^+}{m_1^+},  \qquad \vec{\tau}^+
=\frac{\vec{b}^-}{m_1^+}.
%\end{aligned}
\end{equation*}
There are  some additional relations which ensure that
both $T(\lambda)$ and its inverse $\hat{T}(\lambda)$ belong to the
orthogonal group $SO(2r+1)$ and that $T(\lambda)\hat{T}(\lambda) =\openone$.

Next we introduce the fundamental analytic solution (FAS) $\chi^{\pm}
(x,t,\lambda )$  using  the generalized
Gauss decomposition of $T(\lambda,t)$, see \cite{ZMNP,VSG2,TMF98}:
\begin{equation}\label{eq:FAS_J}
%\begin{split}
\chi ^\pm(x,t,\lambda)= \phi (x,t,\lambda) S_{J}^{\pm}(t,\lambda )
= \psi (x,t,\lambda ) T_{J}^{\mp}(t,\lambda ) D_J^\pm (\lambda),
\end{equation}
This construction ensures that
$\xi^\pm(x,\lambda)=\chi^\pm(x,\lambda) e^{i\lambda Jx}$
are analytic functions of $\lambda$ for $\lambda \in \bbbc_\pm$.
If $Q(x,t) $ is a solution of the MNLS eq. (\ref{eq:4.2}) then  the
matrix  elements of $T(\lambda)$  satisfy the  linear evolution equations \cite{GKV*09}
\begin{equation}\label{eq:evol}
\begin{aligned}
i\frac{d\vec{b}^{\pm}}{d t} \pm \lambda ^2
\vec{b}^{\pm}(t,\lambda ) &=0, & \qquad i\frac{d\vec{B}^{\pm}}{d t}
\pm \lambda ^2 \vec{B}^{\pm}(t,\lambda ) &=0, \\
i\frac{d m_1^{\pm}}{d t}  &=0, &\qquad  i \frac{d{\bf
m}_2^{\pm}}{d t}  &=0.
\end{aligned}
\end{equation}
Thus the block-diagonal matrices $D^{\pm}(\lambda)$ can be
considered as generating functionals of the integrals of motion.
The fact that all $(2r-1)^2$ matrix elements of
$\m_2^\pm(\lambda)$ for $\lambda \in \bbbc_\pm$  generate
integrals of motion reflect the superintegrability of the model
and are due to the degeneracy of the dispersion law of
(\ref{eq:4.2}). Note that $D^\pm_J(\lambda)$ allow analytic
extension for $\lambda\in \bbbc_\pm$ and that their zeroes and
poles determine the discrete eigenvalues\cite{GKKV*08} of $L$.

\subsection{The Riemann-Hilbert Problem}
The FAS for real $\lambda$ are linearly related \cite{GKV*09}
\begin{equation}\label{eq:rhp0}
\chi^+(x,t,\lambda) =\chi^-(x,t,\lambda) G_{0,J}(\lambda,t),
\qquad G_{0,J}(\lambda,t)=\hat{S}^-_J(\lambda,t)S^+_J(\lambda,t)
\end{equation}
Eq. (\ref{eq:rhp0}) can be rewriten in an equivalent form for the
FAS $\xi^\pm(x,t,\lambda)=\chi^\pm (x,t,\lambda)e^{i\lambda Jx }$:
\begin{equation}\label{eq:xi}
i\frac{d\xi^\pm}{dx} + Q(x)\xi^\pm(x,\lambda) -\lambda [J,
\xi^\pm(x,\lambda)]=0,
\end{equation}
and  the relation
\begin{equation}\label{eq:rh-n}
\lim_{\lambda \to \infty} \xi^\pm(x,t,\lambda) = \openone, \qquad
\end{equation}
Then these FAS satisfy the RHP's
\begin{equation}\label{eq:rhp1}
\xi^+(x,t,\lambda) =\xi^-(x,t,\lambda) G_J(x,\lambda,t), \qquad
G_{J}(x,\lambda,t) =e^{-i\lambda J(x+\lambda t)}G^-_{J}(\lambda,t)e^{i\lambda J(x+\lambda t)} ,
\end{equation}
Obviously the sewing function $G_J(x,\lambda,t)$   is uniquely
determined by the Gauss factors $S_J^\pm (\lambda,t)$. In
addition Zakharov-Shabat's theorem \cite{ZaSh1,ZaSh2} states that  $G_J(x,\lambda,t)$  depends
on $x$ and $t$ in the way prescribed above then the corresponding FAS satisfy the
linear systems (\ref{eq:xi}).

If we have solved the RHP's  and know the FAS $\xi^+(x,t,\lambda)$ then the formula
\begin{equation}\label{eq:XI-Q}
    Q(x,t) = \lim_{\lambda\to\infty} \lambda \left( J- \xi^+(x,t,\lambda)J\hat{\xi}^+(x,t,\lambda)
    \right),
\end{equation}
allows us to recover the corresponding potential of $L$.

\section{Reductions of MNLS}

Along with the typical reduction $Q=Q^\dag$ mentioned above one can impose additional reductions
using the reduction group proposed by Mikhailov \cite{Mikh}.
They are automatically compatible with the Lax representation of the corresponding MNLS eq. Below we
make use of two $\bbbz_2$-reductions\cite{R95}:
\begin{equation}\label{eq:U-V.a}
\begin{aligned}
&\mbox{1)} &\qquad C_1 U^{\dagger}(x,t, \lambda^* ) C_1^{-1}&= U(x,t,\lambda ),
&\qquad C_1 V^{\dagger}( x,t,\lambda^* ) C_1^{-1}&= V(x,t,\lambda ), \\
&\mbox{2)}  &\qquad C_2 U^{T}(x,t,\lambda )C_2^{-1} &= -U(x,t,\lambda ), &\qquad
C_2 V^{T}( x,t,\lambda )C_2^{-1} &= -V(x,t,\lambda ),
\end{aligned}
\end{equation}
where $C_1$ and $C_2$ are involutions of the Lie algebra $so(2r+1$, i.e.   $C_i^2=\openone$. They
can be chosen to be either diagonal (i.e., elements of the Cartan subgroup of $SO(2r+1)$) or
elements of the Weyl group.

The typical reductions of the MNLS eqs. is a  class 1) reduction obtained by specifying
$C_1$ to be the identity automorphism of $\fr{g}$; below we list several choices for $C_1$
leading to inequivalent reductions:
\begin{equation}\label{eq:C1-1}
\begin{aligned}
\mbox{1a)} \quad C_1 &=\openone, &\quad \vec{p}(x) &=\vec{q}{\,}^*(x) , \quad
&\mbox{1b)} \quad C_1&=K_1, & \vec{p}(x) &=K_{01}\vec{q}{\,}^*(x) , \\
\mbox{1c)} \quad C_1 &=S_{e_2}, & \quad \vec{p}(x) &=K_{02}\vec{q}{\,}^*(x) , \quad &\mbox{1d)}
\quad C_1&=S_{e_2}S_{e_3},  & \vec{p}(x) &=K_{03}\vec{q}{\,}^*(x) ,\\
\mbox{2e)} \quad C_2 &=K_4, &\quad \vec{q}(x) &=- K_{04} s_0\vec{q}(x) , \quad
&  \quad   \vec{p}(x) &=-K_{04} s_0\vec{p}(x) ,
\end{aligned}
\end{equation}
where
\begin{equation}\label{eq:C1-2}
K_j =\bdiag(1, K_{0j}, 1), \qquad K_{01} = \diag(\epsilon_1,\dots ,\epsilon_{r-1},1, \epsilon_{r-1}, \dots,
\epsilon_1 ), \quad j=1,2,3,
\end{equation}
and $\epsilon_j=\pm 1$.
The matrices $K_{02}$, $K_{03}$ and $K_4$ are not diagonal and may take the form:
\begin{equation}\label{eq:C1-3}
\begin{aligned}
 K_{02} &= \left( \begin{array}{ccc} 0 & 0 & 1 \\ 0 & -1 & 0 \\ 1 & 0 & 0 \end{array}\right), &
 \qquad  K_{4} &= \left( \begin{array}{ccc} 0 & 0 & 1 \\ 0 & K_{04} & 0 \\ 1 & 0 & 0 \end{array}\right), & \\
 K_{02} &= \left( \begin{array}{ccccc} 0 & 0 & 0 & 0 & -1 \\ 0 & 1 &0 & 0 & 0 \\
0 & 0 & -1 & 0 & 0 \\ 0 & 0 & 0 & 1 & 0  \\ -1 & 0 & 0 & 0 & 0  \end{array}\right), & \qquad
K_{03} &= \left( \begin{array}{ccccc} 0 & 0 & 0 & 0 & -1 \\ 0 & 0 &0 & 1 & 0 \\
0 & 0 & -1 & 0 & 0 \\ 0 & 1 & 0 & 0 & 0  \\ -1 & 0 & 0 & 0 & 0  \end{array}\right), &
\end{aligned}
\end{equation}

Each of the above reductions impose constraints on the FAS, on the scattering matrix $T(\lambda)$ and on its
Gauss factors $S^\pm_J(\lambda)$, $T^\pm_J(\lambda)$ and $D^\pm_J(\lambda)$. These have the form:
\begin{equation}\label{eq:C1-3'}
\begin{aligned}
(S^+(\lambda^*))^\dag &= K_j^{-1}\hat{S}^-(\lambda)K_j  &\quad (T^+(\lambda^*))^\dag &=K_j^{-1}\hat{T}^-(\lambda)K_j  &\quad
(D^+(\lambda^*))^\dag &= K_j^{-1}\hat{D}^-(\lambda)K_j   \\
\vec{\tau}{\,}^+ &= K_{0j}\vec{\tau}{\,}^{-,*}, &\quad  \vec{\rho}{\,}^+ &= K_{0j}\vec{\rho}{\,}^{-,*}, &
\end{aligned}
\end{equation}
where the matrices $K_j$ are specific for each choice of the automorphisms $C_1$, see eqs. (\ref{eq:C1-1}), (\ref{eq:C1-2}).

In particular, from the last line of (\ref{eq:C1-3'}) and (\ref{eq:C1-2}) we get:
\begin{equation}\label{eq:m1pm}
(m_1^+(\lambda^*))^* = m_1^-(\lambda),
\end{equation}
and consequently, if $m_1^+(\lambda)$ has zeroes at the points $\lambda_k^+$, then
$m_1^-(\lambda)$ has  zeroes at:
\begin{equation}\label{eq:lapm}
\lambda_k^- = (\lambda_k^+)^*, \qquad k=1,\dots, N.
\end{equation}

\section{Soliton solutions}

Let us now make use of one of the   versions of the dressing method\cite{ZaSh1,ZaSh2} which allows one to
construct singular solutions of the RHP. In order to obtain $N$-soliton solutions one has to apply dressing procedure with
a $2N$-poles dressing factor of the form
\begin{equation}
u(x,\lambda)=\openone+\sum^N_{k=1}\left(\frac{A_k(x)}{\lambda-\lambda^+_k}
+\frac{B_k(x)}{\lambda-\lambda^-_k}\right).	
\label{dressfac_bcd_n}\end{equation}
The $N$-soliton solution itself can be generated via the following formula
\begin{equation}
Q_{N,\rm s}(x)= \sum^N_{k=1}[J,A_k(x)+B_k(x)].	
\label{dressq_bcd_n}\end{equation}
The dressing
factor $u(x,\lambda ) $ must satisfy the equation
\begin{equation}\label{eq:u-eq}
i\partial_x u + Q_{N,\rm s} u -  \lambda [J,u] =0
\end{equation}
and the normalization condition $\lim_{\lambda \to\infty }
u(x,\lambda ) =\openone $. The construction of $u(x,\lambda )\in
SO(n+2) $ is based on an appropriate anzatz specifying the form
of its $\lambda $-dependence \cite{ZaMi,GGK05a}

The residues of $u$ admit the following decomposition
\[A_k(x)=X_k(x)F^T_k(x),\qquad B_k(x)=Y_k(x){G}^T_k(x).\]
where all matrices involved are supposed to be rectangular and of maximal rank $s$.
By comparing the coefficients before the same powers of $\lambda-\lambda^{\pm}_k$
in (\ref{eq:u-eq}) we convince ourselves that the factors $F_k$ and $G_k$ can be
expressed by the fundamental analytic solutions $\chi^{\pm}_0(x,\lambda)$ as follows
\[F^T_k(x)=F^T_{k,0}[\chi^{+}_0(x,\lambda^+_k)]^{-1},\qquad G^T_k(x)=G^T_{k,0}[\chi^{-}_0(x,\lambda^-_k)]^{-1}.\]
The constant rectangular matrices $F_{k,0}$ and $G_{k,0}$ obey the algebraic relations
\[F^T_{k,0}S_0F_{k,0}= 0,\qquad G^T_{k,0}S_0G_{k,0}=0.\]

The other two types of factors $X_k$ and $Y_k$ are solutions to the algebraic system
\begin{equation}\label{eq:XY}
\begin{split}
S_0F_k &=X_k\alpha_k+\sum_{l\neq k}\frac{X_lF^T_lS_0F_k}{\lambda^+_l-\lambda^+_k}
+\sum_l\frac{Y_lG^T_lS_0F_k}{\lambda^-_l-\lambda^+_k},\\
S_0G_k &=\sum_{l}\frac{X_lF^T_lS_0G_k}{\lambda^+_l-\lambda^-_k}+Y_k\beta_k
+\sum_{l\neq k}\frac{Y_lG^T_lS_0G_k}{\lambda^-_l-\lambda^-_k}.
\end{split}
\end{equation}
The square $s\times s$ matrices $\alpha_k(x)$ and $\beta_k(x)$ introduced above depend on
$\chi^+_0$ and $\chi^-_0$ and their derivatives by $\lambda$ as follows
\begin{equation}\label{eq:XY-alf}
\begin{split}
\alpha_k(x)&=-F^T_{0,k}[\chi^{+}_0(x,\lambda^{+}_k)]^{-1}
\partial_{\lambda}\chi^{+}_0(x,\lambda^{+}_k)S_0F_{0,k}+\alpha_{0,k},\\
\beta_k(x)&=-G^T_{0,k}[\chi^{-}_0(x,\lambda^{-}_k)]^{-1}
\partial_{\lambda}\chi^{-}_0(x,\lambda^{-}_k)S_0G_{0,k}+\beta_{0,k}.
\end{split}
\end{equation}

Below for simplicity we will choose $F_k$ and $G_k$ to be $2r+1$-component vectors. Then one can show %\cite{Valch}
that $\alpha_k=\beta_k=0$ which simplifies the system (\ref{eq:XY}). We also introduce the following
more convenient parametrization for $F_k$ and $G_k$, namely (see eq. (\ref{eq:njxt})):
\begin{equation}\label{eq:FkGk}
F_k(x,t) = S_0|n_k(x,t)\rangle = \left( \begin{array}{c} e^{-z_k+i\phi_k} \\
-\sqrt{2} s_0 \vec{\nu}_{0k} \\ e^{z_k-i\phi_k} \end{array} \right), \qquad
G_k(x,t) = |n_k^*(x,t)\rangle = \left( \begin{array}{c} e^{z_k+i\phi_k} \\
\sqrt{2} \vec{\nu}_{0k}{\,}^* \\ e^{-z_k-i\phi_k} \end{array} \right),
\end{equation}
where $\vec{\nu}_{0k}$ are constant $2r-1$-component polarization vectors and
\begin{equation}\label{eq:njxt}
\begin{split}
z_j = \nu_j(x+2\mu_j t) + \xi_{00} , &\qquad \phi_j = \mu_j x+(\mu_j^2-\nu_j^2) t +\delta_{00}, \\
\langle n_j^T(x,t)| S_0|n_j(x,t)\rangle =0, \qquad &\mbox{or} \qquad (\vec{\nu}_{0,j} s_0 \vec{\nu}_{0,j}) = 1.
\end{split}
\end{equation}
With this notations the polarization vectors automatically satisfy $\langle n_j (x,t) |S_0| n_j(x,t)\rangle =0$.

Thus for $N=1$ we get the system:
\begin{equation}\label{eq:N1}
%\begin{split}
|Y_1\rangle = -\frac{ (\lambda_1^+ -\lambda_1^-) |n_1\rangle}{\langle n_1^\dag | n_1\rangle} , \qquad
|X_1\rangle = \frac{ (\lambda_1^+ -\lambda_1^-) S_0 |n_1^*\rangle}{\langle n_1^\dag | n_1\rangle} ,
%\end{split}
\end{equation}
which is easily solved. As a result for the one-soliton solution we get:
\begin{equation}\label{eq:1s}
\vec{q}_{\rm 1s} = -\frac{i \sqrt{2}(\lambda_1^+ -\lambda_1^-) e^{-i\phi_1} }{\Delta_1}
\left( e^{-z_1} s_0|\vec{\nu}_{01}\rangle + e^{z_1} |\vec{\nu}_{01}^*\rangle \right), \qquad \Delta_1 =
\cosh (2z_1) +  \langle\vec{\nu}_{01}^\dag   |\vec{\nu}_{01}\rangle .
\end{equation}

For $n=3$ we put $\nu_{0k} = |\nu_{0k}|e^{\alpha_{0k}}$ get:
\begin{equation}\label{eq:1s-1}
\begin{split}
\Phi_{{\rm 1s};\pm 1} &= -\frac{\sqrt{2 |\nu_{01;1}\nu_{01;3}|}(\lambda_1^+ -\lambda_1^-)  }{\Delta_1} e^{-i\phi_1 \pm i\beta_{13}}
\left( \cosh(z_1 \mp \zeta_{01}) \cos(\alpha_{13}) - i\sinh(z_1 \mp \zeta_{01}) \sin(\alpha_{13})   \right), \\
\Phi_{{\rm 1s};0} &= -\frac{\sqrt{2} |\nu_{01;2}|(\lambda_1^+ -\lambda_1^-)  }{\Delta_1} e^{-i\phi_1 }
\left( \sinh{z_1 } \cos(\alpha_{02}) +i\cosh{z_1 } \sin(\alpha_{02})   \right), \\
\beta_{13} &= \frac{1}{2} (\alpha_{03}-\alpha_{01}) , \qquad \zeta_{01} = \frac{1}{2} \ln  \frac{|\nu_{01;3}|}{|\nu_{01;1}|},
\qquad \alpha_{13} = \frac{1}{2} (\alpha_{03}+\alpha_{01}) ,
\end{split}
\end{equation}
Note that the `center of mass` of $\Phi_{{\rm 1s};1}$  (resp. of $\Phi_{{\rm 1s};-1}$) is shifted with respect to the one of
$\Phi_{{\rm 1s};0}$ by $\zeta_{01}$ to the right (resp to the left); besides $|\Phi_{{\rm 1s};1}|=|\Phi_{{\rm 1s};-1}|$, i.e. they have
the same amplitudes.

For $n=5$ we put $\nu_{0k} = |\nu_{0k}|e^{\alpha_{0k}}$ and get analogously:
\begin{equation}\label{eq:1s-3}
\begin{split}
\Phi_{{\rm 1s};\pm 2} &= -\frac{\sqrt{2 |\nu_{01;1}\nu_{01;5}|}(\lambda_1^+ -\lambda_1^-)  }{\Delta_1} e^{-i\phi_1\pm i\beta_{15}}
\left( \cosh(z_1\mp \zeta_{01}) \cos(\alpha_{15}) - i\sinh(z_1\mp \zeta_{01}) \sin(\alpha_{15})   \right), \\
\Phi_{{\rm 1s};\pm 1} &= \frac{\sqrt{2 |\nu_{01;2}\nu_{01;4}|}(\lambda_1^+ -\lambda_1^-)  }{\Delta_1} e^{-i\phi_1\pm i\beta_{24}}
\left( \cosh(z_1\mp \zeta_{02}) \cos(\alpha_{24}) -i\sinh(z_1\mp \zeta_{01}) \sin(\alpha_{24})   \right),\\
\Phi_{{\rm 1s};0} &= -\frac{\sqrt{2} |\nu_{01;3}|(\lambda_1^+ -\lambda_1^-)  }{\Delta_1} e^{-i\phi_1 }
\left( \cosh{z_1 } \cos(\alpha_{03}) -i\sinh{z_1 } \sin(\alpha_{03})   \right), \\
\beta_{15} &= \frac{1}{2} (\alpha_{05}-\alpha_{01}) , \qquad \zeta_{01} = \frac{1}{2} \ln  \frac{|\nu_{01;5}|}{|\nu_{01;1}|},
\qquad \alpha_{15} = \frac{1}{2} (\alpha_{05}+\alpha_{01}) , \\
\beta_{24} &= \frac{1}{2} (\alpha_{04}-\alpha_{02}) , \qquad \zeta_{02} = \frac{1}{2} \ln  \frac{|\nu_{01;4}|}{|\nu_{01;2}|},
\qquad \alpha_{24} = \frac{1}{2} (\alpha_{04}+\alpha_{02}) ,
\end{split}
\end{equation}
Similarly the `center of mass` of $\Phi_{{\rm 1s};2}$  and  $\Phi_{{\rm 1s};1}$  (resp. of $\Phi_{{\rm 1s};-2}$ and
$\Phi_{{\rm 1s};-1}$) are shifted with respect to the one of $\Phi_{{\rm 1s};0}$ by $\zeta_{01}$  and $\zeta_{02}$
to the right (resp to the left); besides $|\Phi_{{\rm 1s};2}|=|\Phi_{{\rm 1s};-2}|$ and $|\Phi_{{\rm 1s};1}|=|\Phi_{{\rm 1s};-1}|$.

For $N=2$ we get:
\begin{equation}\label{eq:n-nst}
\begin{split}
|n_1(x,t)\rangle & = \frac{X_2(x,t) f_{21}}{\lambda_2^+ -\lambda_1^+} +
\frac{Y_1(x,t) \kappa_{11}}{\lambda_1^- -\lambda_1^+} + \frac{Y_2(x,t) \kappa_{21}}{\lambda_2^- -\lambda_1^+}, \\
|n_2(x,t)\rangle & = \frac{X_1(x,t) f_{12}}{\lambda_1^+ -\lambda_2^+} +
\frac{Y_1(x,t) \kappa_{12}}{\lambda_1^- -\lambda_2^+} + \frac{Y_2(x,t) \kappa_{22}}{\lambda_2^- -\lambda_2^+}, \\
S_0|n_1^*(x,t)\rangle & = \frac{X_1(x,t) \kappa_{11}}{\lambda_2^+ -\lambda_1^+} +
\frac{X_2(x,t) \kappa_{11}}{\lambda_2^+ -\lambda_1^-} + \frac{Y_2(x,t) f_{21}^*}{\lambda_2^- -\lambda_1^-}, \\
S_0|n_2^*(x,t)\rangle & = \frac{X_1(x,t) \kappa_{21}}{\lambda_1^+ -\lambda_2^-} +
\frac{X_2(x,t) \kappa_{22}}{\lambda_2^+ -\lambda_2^-} + \frac{Y_1(x,t) f_{12}^*}{\lambda_1^- -\lambda_2^-},
\end{split}
\end{equation}
where
\begin{equation}\label{eq:n-nst2}
\begin{split}
\kappa_{kj}(x,t)& =  e^{z_k+z_j+i(\phi_k-\phi_j)} +e^{-z_k-z_j-i(\phi_k-\phi_j)}
+ 2 \left( \vec{\nu}{\,}^\dag_{0k},\vec{\nu}_{0j}\right) ,\\
f_{kj}(x,t) &= e^{z_k-z_j-i(\phi_k-\phi_j)} +e^{z_j-z_k+i(\phi_k-\phi_j)}
- 2 \left( \vec{\nu}_{0k}^T s_0\vec{\nu}_{0j}\right) ,
\end{split}
\end{equation}

In other words:
\begin{equation}\label{eq:MX-n}
\mathcal{M} \vec{X} \equiv \left( \begin{array}{cccc} 0 & \frac{f_{21}}{\lambda_2^+ -\lambda_1^+} &
\frac{\kappa_{11}}{\lambda_1^- -\lambda_1^+} & \frac{\kappa_{21}}{\lambda_2^- -\lambda_1^+} \\
\frac{f_{12}}{\lambda_1^+ -\lambda_2^+} & 0 & \frac{\kappa_{12}}{\lambda_1^- -\lambda_2^+} &
\frac{\kappa_{22}}{\lambda_2^- -\lambda_2^+} \\
\frac{\kappa_{11}}{\lambda_1^+ -\lambda_1^-} & \frac{\kappa_{12}}{\lambda_2^+ -\lambda_1^-} & 0
& \frac{f_{21}^*}{\lambda_2^- -\lambda_1^-}  \\
\frac{\kappa_{21}}{\lambda_1^+ -\lambda_2^-} & \frac{\kappa_{22}}{\lambda_2^+ -\lambda_2^-}
& \frac{f_{12}^*}{\lambda_1^- -\lambda_2^-} & 0 \\ \end{array} \right) \left( \begin{array}{c} X_1 \\ X_2 \\
Y_1 \\ Y_2 \\ \end{array} \right) = \left( \begin{array}{c} |n_1\rangle \\ |n_2\rangle  \\
S_0|n_1^*\rangle  \\ S_0|n_2^*\rangle \\ \end{array} \right) .
\end{equation}
We can rewrite $\mathcal{M}$ in block-matrix form:
\begin{equation}\label{eq:MM}
\begin{split}
\mathcal{M} &= \left( \begin{array}{cc} \mathcal{M}_{11} & \mathcal{M}_{12} \\ \mathcal{M}_{21} & \mathcal{M}_{22} \end{array}
\right) , \qquad \mathcal{M}_{22} = \mathcal{M}_{11}^*, \qquad  \mathcal{M}_{21} = -\mathcal{M}_{12}^T, \\
\mathcal{M}_{11} &=  \frac{f_{12}}{\lambda_2^+ -\lambda_1^+} \left( \begin{array}{cc} 0 & 1 \\ -1 & 0 \\ \end{array} \right),
\qquad   \mathcal{M}_{12} =    \left( \begin{array}{cc} \frac{\kappa_{11}}{\lambda_1^- -\lambda_1^+} &
\frac{\kappa_{21}}{\lambda_2^- -\lambda_1^+} \\ \frac{\kappa_{12}}{\lambda_1^- -\lambda_2^+} & \frac{\kappa_{22}}{\lambda_2^- -\lambda_2^+} \\ \end{array} \right).
\end{split}
\end{equation}
The inverse of $ \mathcal{M}  $ is given by:
\begin{equation}\label{eq:MM-1}
\mathcal{M}^{-1} = \left( \begin{array}{cc}
(\mathcal{M}_{11} - \mathcal{M}_{12}\hat{\mathcal{M}}_{11}^* \mathcal{M}_{21})^{-1} &
-(\mathcal{M}_{11} - \mathcal{M}_{12}\hat{\mathcal{M}}_{11}^* \mathcal{M}_{21})^{-1}\mathcal{M}_{12} \hat{\mathcal{M}}_{11}^*\\
-(\mathcal{M}_{11}^* - \mathcal{M}_{21}\hat{\mathcal{M}}_{11} \mathcal{M}_{12})^{-1}  \mathcal{M}_{21}  \hat{\mathcal{M}}_{11}
& (\mathcal{M}_{11}^* - \mathcal{M}_{21}\hat{\mathcal{M}}_{11} \mathcal{M}_{12})^{-1} \end{array}
\right) , \qquad
\end{equation}
One can check by direct calculation that:
\begin{equation}\label{eq:MM1}
\begin{split}
\mathcal{M}_{11} -\mathcal{M}_{12}\hat{\mathcal{M}}_{11}^*\mathcal{M}_{21} =
\frac{f_{12}^*} {\lambda_2^- - \lambda_1^-} Z \left( \begin{array}{cc} 0 & 1 \\
-1 & 0 \\ \end{array} \right), \\
\mathcal{M}_{11}^* -\mathcal{M}_{21}\hat{\mathcal{M}}_{11}\mathcal{M}_{12} =
\frac{f_{12}}{\lambda_2^+ - \lambda_1^+} Z \left( \begin{array}{cc} 0 & 1 \\
-1 & 0 \\ \end{array} \right), \\
Z=  \left( \frac{|f_{12}|^2}{|\lambda_2^+-\lambda_1^+|^2} -
\frac{\kappa_{12}\kappa_{21}}{|\lambda_2^+-\lambda_1^-|^2}  +\frac{\kappa_{11}\kappa_{22}}{4\nu_1\nu_2} \right),
\end{split}
\end{equation}
Finally we get:
\begin{equation}\label{eq:M-1}
\mathcal{M}^{-1} = \frac{1}{Z} \left( \begin{array}{cccc} 0 & \frac{f_{12}^*}{\lambda_1^- -\lambda_2^-} &
-\frac{\kappa_{22}}{\lambda_2^+ -\lambda_2^-} & \frac{\kappa_{12}}{\lambda_2^+ -\lambda_1^-} \\
-\frac{f_{12}^*}{\lambda_1^- -\lambda_2^-} & 0 & \frac{\kappa_{21}}{\lambda_1^+ -\lambda_2^-} & -\frac{\kappa_{11}}{\lambda_1^+ -\lambda_1^-} \\
\frac{\kappa_{22}}{\lambda_2^+ -\lambda_2^-} & -\frac{\kappa_{21}}{\lambda_1^+ -\lambda_2^-} & 0 & -\frac{f_{12}}{\lambda_1^+ -\lambda_2^+}  \\
-\frac{\kappa_{12}}{\lambda_2^+ -\lambda_1^-} & \frac{\kappa_{11}}{\lambda_1^+ -\lambda_1^-}  & \frac{f_{12}}{\lambda_2^+ -\lambda_1^+} & 0 \\
\end{array} \right) ,
\end{equation}

From eqs. (\ref{eq:MX-n}) and (\ref{eq:M-1}) we obtain:
\begin{equation}\label{eq:MM1'}
\begin{split}
|X_1\rangle &= \frac{1}{Z} \left( \frac{f_{12}^*}{\lambda_1^- -\lambda_2^-} |n_2\rangle -
\frac{\kappa_{22}}{\lambda_2^+ -\lambda_2^-} S_0 |n_1^*\rangle + \frac{\kappa_{12}}{\lambda_2^+
-\lambda_1^-} S_0 |n_2^*\rangle  \right), \\
|X_2\rangle &= \frac{1}{Z} \left( -\frac{f_{12}^*}{\lambda_1^- -\lambda_2^-} |n_1\rangle +
\frac{\kappa_{21}}{\lambda_1^+ -\lambda_2^-} S_0 |n_1^*\rangle - \frac{\kappa_{11}}{\lambda_1^+
-\lambda_1^-} S_0 |n_2^*\rangle  \right), \\
|Y_1\rangle &= \frac{1}{Z} \left( \frac{\kappa_{22}}{\lambda_2^+ -\lambda_2^-} |n_1\rangle -
\frac{\kappa_{21}}{\lambda_1^+ -\lambda_2^-} |n_2\rangle - \frac{f_{12}}{\lambda_1^+
-\lambda_2^+} S_0 |n_2^*\rangle  \right), \\
|Y_2\rangle &= \frac{1}{Z} \left( -\frac{\kappa_{12}}{\lambda_2^+ -\lambda_1^-} |n_1\rangle +
\frac{\kappa_{11}}{\lambda_1^+ -\lambda_1^-} |n_2\rangle + \frac{f_{12}}{\lambda_2^+
-\lambda_1^+} S_0 |n_1^*\rangle  \right),
\end{split}
\end{equation}

Inserting this result into eq. (\ref{dressq_bcd_n}) we obtain the following expression for the 2-soliton
solution of the MNLS:
\begin{equation}\label{eq:2-sol}
\begin{split}
Q_{\rm 2s}(x,t) &= [J, A_1+B_1+A_2+B_2] = \frac{1}{Z} [J , C(x,t) - S_0C^T(x,t) S_0], \\
C(x,t) &=  \frac{\kappa_{22}}{\lambda_2^+ -\lambda_2^-} |n_1\rangle  \langle n_1^\dag | - \frac{\kappa_{12}}{\lambda_2^+ -\lambda_1^-}
|n_1\rangle \langle n_2^\dag | - \frac{\kappa_{21}}{\lambda_1^+ -\lambda_2^-} |n_2\rangle  \langle n_1^\dag | + \frac{\kappa_{11}}{\lambda_1^+ -\lambda_1^-} |n_2\rangle  \langle n_2^\dag |  \\ & -
\frac{f_{12}^*}{\lambda_1^- -\lambda_2^-} |n_1\rangle \langle n_2| S_0 - \frac{f_{12}}{\lambda_1^+ -\lambda_2^+}
S_0|n_2^*\rangle  \langle n_1^\dag | .
\end{split}
\end{equation}

At the end of this section we note that the effect of the reductions (\ref{eq:U-V.a})--(\ref{eq:C1-1})
consists in constraining the polarization vectors. For the reduction 2e) we get
\begin{equation}\label{eq:red-nu}
\vec{\nu}_{0k}= K_{04}\vec{\nu}_{0k}
\end{equation}
In particular, for $n=3$ and for $K_{04}= - \openone $ we have $ q_1=q_3$, and $q_2$  arbitrary.
This reduction of eq. (\ref{eq:1}) is also important for the BEC \cite{Kevre*08}. From
(\ref{eq:red-nu}) we find $\nu_{01}=\nu_{03}$. The effect of this constraint is that for the one-soliton
solution we get $\Phi_{{\rm 1s};1} =\Phi_{{\rm 1s};3}$.

Our next remark following  \cite{ps99} is that this reduction applied to the $F=1$ MNLS (\ref{eq:1}) leads to a 2-component MNLS
which after the change of variables
\begin{equation}\label{eq:2MNLS}
q_1 = \frac{1}{2} (w_1 + iw_2), \qquad q_2 =  \frac{1}{\sqrt{2}}  (w_1 - iw_2),
\end{equation}
leads to two disjoint NLS equations for $w_1$ and $w_2$ respectively.

It is only logical that applying the constraint $\nu_{01}= \nu_{03}$ the explicit expression for the one-soliton
solution (\ref{eq:1s-1}) simplifies and reduces to the standard soliton solutions of the scalar NLS.

\section{Two Soliton interactions}

In this section we generalize the classical results of Zakharov and Shabat about soliton interactions \cite{zs72}  to the class
of MNLS equations related to BD.I symmetric spaces. For detailed exposition see the monographs \cite{ZMNP,FaTa}. These results were generalized
for the vector nonlinear Schr\"odinger equation by Manakov \cite{ma74}, see also \cite{AblPrinTru*04,Laksh,WadTsu}.
The Zakharov Shabat approach consisted in calculating the asymptotics of generic $N$-soliton solution of NLS for $t \to \pm\infty$ and establishing
the pure elastic character of the generic soliton interactions. By generic here we mean $N$-soliton solution whose parameters
$\lambda_k^\pm = \mu_k \pm i\nu_k$ are such that $\mu_k \neq \mu_j$ for $k\neq j$. The pure elastic character of the soliton interactions
is demonstrated by the fact that for $t\to\pm\infty$ the generic $N$-soliton solution splits into sum of $N$ one soliton solutions each
preserving  its amplitude $2\nu_k$ and velocity $\mu_k$. The only effect of the interaction consists in shifting the center of mass
and the initial phase of the solitons. These shifts can be expressed in terms of $\lambda_k^\pm$ only; for detailed exposition see \cite{FaTa}.

We start with the simplest non-trivial case. Namely we  use the $2$-soliton solution derived above and
calculate its asymptotics   along the trajectory of the first soliton. To this end
we keep $z_1(x,t)$ fixed and let $\tau =z_2 -z_1$ tend to $\pm \infty$. Therefore it will  be enough to insert the asymptotic values of
the matrix elements of $\mathcal{M}$ for $\tau\to\pm\infty$ and keep only the leading terms. That gives:
\begin{equation}\label{eq:as1}
\begin{split}
\kappa_{22} &= \left\{ \begin{array}{ll} e^{2\tau} \exp(\nu_2z_1/\nu_1) + 2 \mathcal{C}_1 , & \quad \mbox{for} \quad \tau \to \infty, \\
e^{-2\tau} \exp(-\nu_2z_1/\nu_1) + 2 \mathcal{C}_1 , & \quad \mbox{for} \quad \tau \to -\infty, \end{array} \right. \\
\kappa_{12} &= \left\{ \begin{array}{ll} e^{\tau} \exp((1+\nu_2/\nu_1)z_1 +i(\phi_1 -\phi_2)) + \mathcal{O}(1) , & \quad
\mbox{for} \quad \tau \to \infty, \\
 e^{-\tau} \exp(-(1+\nu_2/\nu_1)z_1 -i(\phi_1 -\phi_2)) + \mathcal{O}(1) , & \quad \mbox{for} \quad \tau \to -\infty, \end{array} \right. \\
\kappa_{21} &= \left\{ \begin{array}{ll} e^{\tau} \exp((1+\nu_2/\nu_1)z_1 -i(\phi_1 -\phi_2)) + \mathcal{O}(1) , & \quad
\mbox{for} \quad \tau \to \infty, \\
 e^{-\tau} \exp(-(1+\nu_2/\nu_1)z_1 +i(\phi_1 -\phi_2)) + \mathcal{O}(1) , & \quad \mbox{for} \quad \tau \to -\infty, \end{array} \right. \\
f_{12} &= \left\{ \begin{array}{ll} e^{\tau} \exp(-(1-\nu_2/\nu_1)z_1 +i(\phi_1 -\phi_2)) + \mathcal{O}(1) , & \quad
\mbox{for} \quad \tau \to \infty, \\
 e^{-\tau} \exp((1-\nu_2/\nu_1)z_1 -i(\phi_1 -\phi_2)) + \mathcal{O}(1) , & \quad \mbox{for} \quad \tau \to -\infty, \end{array} \right.
\end{split}
\end{equation}

After somewhat lengthy  calculations we get:
\begin{equation}\label{eq:Zpm}
\begin{split}
\lim_{\tau\to\infty} \vec{q}_{\rm 2s}(x,t) &=-\frac{i\sqrt{2} \nu_1 e^{-i(\phi_1 -\alpha_+)} \left( e^{-z_1-r_+} s_0|\vec{\nu}_{01}\rangle
+  e^{z_1+r_+} |\vec{\nu}_{01}^*\rangle \right) }{\cosh(2(z_1+ r_+)) + (\vec{\nu}_{01}^\dag,\vec{\nu}_{01})},  \\
\lim_{\tau\to -\infty} \vec{q}_{\rm 2s}(x,t) &=\frac{i\sqrt{2} \nu_1 e^{-i(\phi_1 +\alpha_+)} \left( e^{-z_1+r_+} s_0|\vec{\nu}_{01}\rangle
+  e^{z_1-r_+} |\vec{\nu}_{01}^*\rangle \right) }{\cosh(2(z_1- r_+)) + (\vec{\nu}_{01}^\dag,\vec{\nu}_{01})},
\end{split}
\end{equation}
where
\[ r_+ =  \ln \left| \frac{\lambda_1^+ -\lambda_2^+}{\lambda_1^+ -\lambda_2^-} \right|, \qquad
\alpha_+ = \arg \frac{\lambda_1^+ -\lambda_2^+}{\lambda_1^+ -\lambda_2^-}.  \]

In other words the $2$-soliton interaction for the MNLS eqs. related to the BD.I symmetric spaces is the same as the one
of the scalar NLS. Again we have that for large times the 2-soliton solution splits into sum of 1-soliton solutions with shifted
center of masses and phases and the value of these shifts $r_+$ and $\alpha_+$ are independent on the number of components of MNLS.
It will be interesting to check whether the $N$-soliton interactions consist of sequence of elementary 2-soliton
interactions and the shifts are additive.

\section{Conclusions and discussion}

Using the Zakharov-Shabat dressing method we have obtained the two-soliton solution and
have used it to analyze the soliton interactions of the MNLS equation. The conclusion is that
after the interactions the solitons recover their polarization vectors $\nu_{0k}$, velocities
and frequency velocities. The effect of the interaction  is, like in for the scalar NLS equation,
shift of the center of mass $z_1\to z_1 +r_+$ and shift of the phase $\phi_1 \to \phi_1 + \alpha_+$.
Both shifts are expressed through the related eigenvalues $\lambda_j^\pm$ only.

The next step would be to analyze multi-soliton interactions. Our hypothesis is that each soliton
will acquire a total shift of the center of mass that is sum of all elementary shifts from each
two soliton interactions. Similar result is expected for the total phase shift of the soliton.
Proofs of these facts will be published elsewhere.

\end{document}